\documentclass[twocolumn,showpacs,floats]{revtex4}

\usepackage{graphics,graphicx,epsfig}
\usepackage{bm}

\begin{document}

\title{Effect of Vision Angle on the Phase Transition in a Flocking Behavior of Animal Groups}
\author{P. The Nguyen$^{a}$, Sang-Hee Lee$^{b}$ and V. Thanh Ngo$^{c}$\footnote{Corresponding author, e-mail: nvthanh@iop.vast.ac.vn }}
\address{$^{a}$Department of  Natural Science, Duytan University,\\  K7/25 Quang Trung, Haichau, Danang, Vietnam\\
$^{b}$Division of Fusion and Convergence of Mathematical Sciences,\\
 National Institute for Mathematical Sciences, Daejeon, Republic of Korea.\\
$^{c}$Institute of Physics - Vietnam Academy of Science and Technology,\\ 10 Dao Tan, Ngoc Khanh, Ba Dinh, Hanoi, Vietnam.
}

\begin{abstract}
The nature of the phase transition in a system of self-propelling particles has been extensively studied during the last few decades. A theoretical model was proposed by T. Vicsek, {\it et. al.} [Phys. Rev. Lett. {\bf 75}, 1226 (1995)] with a simple rule for updating the direction of motion of each particle. Based on the Vicsek's model (VM)~\cite{Vicsek95}, in this work, we consider a group of animals as particles moving freely on a two-dimensional space. Due to the fact that the viewable area of animals depends on the species, we consider the motion of each individual within an angle $\varphi=\phi/2$ ($\phi$ is called angle of view) of a circle centered at its position, of radius $R$.  We obtained a phase diagram in the space ($\varphi$, $\eta_c$) with $\eta_c$ being the critical noise.  We show that, the phase transition exists only in the case of a wide view's angle $\varphi \geq 0.5\pi$. The flocking of animals is an universal behavior of the species of prey, but not the one of the predator. Our simulation results are in good agreement with experimental observation~\cite{Beccoa}.
\end{abstract}

\maketitle

\section{Introduction}

One of the most familiar and intriguing examples of non-equilibrium dynamical systems with many degrees of freedom is a flocking behavior which has been a phenomenon of long standing interest. Well-known examples are found in populations such as large schools of fish,  gathering of birds, swarming of ants and herding of sheep~\cite{Beccoa,Huth,Maldonado,Furbay,Hamilton,Couzin}. Biologically, it has been known that the flocking behavior is advantageous for survival of a population~\cite{Werner,Pitcher,Cresswell}: reducing the risk of capture by predator, higher mating efficiency, easier search for food, efficient learning of external stimuli, and reducing overall aggression~\cite{Cashing,Parrish,Adioui,Zheng}.

In 1987, Reynolds first suggested a simple model consisting of three rules: separation, alignment, and cohesion rules~\cite{Reynolds,Partridge}. These rules describe the behavior of each individual in interaction with other neighboring individuals. The separation rule represents the avoiding behavior among crowding neighbors, the alignment rule describes the steering behavior of individuals towards average heading of neighbors, and the cohesion rule expresses the steering behavior of individuals towards the average position of neighbors. The goal of the model is to generate realistic looking bird flocks in computer animations. All or some of the three rules were mathematically expressed and then analyzed by Vicsek and his coworkers~\cite{Vicsek95,vicsek00,vicsek12}. They mainly focused on the transition between coherently moving and runaway in a stampede.

So far, the flocking behavior has been conventionally studied through simulation in two frameworks: population (Eulerian or continuum models) and individual (agents or particle-based models)~\cite{Adioui,Grunbaum,Inada}. In the population framework, the flock was collectively addressed while flock-density was used as a key variable to present spatial and temporal dynamics of aggregation frequently with partial differential equations of advection-diffusion reaction~\cite{Mogilner,Murray}.
In the individual framework, the flock of agents has been simulated by using ordinary and stochastic equations of motion to describe interactions among agents~\cite{Vicsek95,Huth,Niwa,Tu,Toner}. This approach attempted to replicate naturally observed phenomena from not only animal groups but also other self-propelled characteristics~\cite{Spector} and to compare the evolved characteristics with those of actual animal flocking~\cite{Wood} in order to better understand the possible mechanisms by which these characteristics may have evolved.

Recently, many studies have been made about the effect of vision on the dynamics of flocking behavior~\cite{Lebar,Jure,Ballerini08,MBallerini08,Kunz12,Olson,Strandburg}. Bajec, {\it et. al.}~\cite{Lebar,Jure} proposed a mathematical model which is based on fuzzy logic, so called the fuzzy individual based model. The rules of interactions among the individuals have been suggested that each individual is only under the influence of around seven nearest neighbors~\cite{Ballerini08,MBallerini08}, or to be constant across group sizes~\cite{Kunz12}. Using an evolutionary model of a predator-prey system, one showed that predator confusion helps to evolve swarming behaviour in prey~\cite{Olson}.  Strandburg-Peshkin, {\it et. al.}~\cite{Strandburg} showed that visual interaction networks are specified by each model which depend on the number of nearest neighbors, interaction radius,
or visual threshold.

In the present study, we consider a population consisting of identical individuals with a biological property: the vision angle. Most animals are able to distinguish the predator using their vision capability. We explored the phase transition from order to disorder in movement of individuals and briefly discussed the biological meaning for the change of the phase transition in accordance with the vision angle.

\section{The Model}
The theoretical model proposed by Vicsek {\it et al.}~\cite{Vicsek95}, consists of $N$ individuals, labeled by an index $i$ ranging from $1$ to $N$, continuously moving in a plane ($x$, $y$) of linear size $L$. The $i$th individual is characterized by their position $\bm r_i = (x_i, y_i)$ and velocity $\bm v_i$ of fixed modulus $\left| \bm v_i\right| = v_0$. Periodic boundary conditions are used in the two directions.

In the simulations, we use the following initial conditions: at time $t = 0$, the individuals are randomly distributed in the plane with the same constant velocity modulus $v_0$. For the orientation of each individual, we generate an angle $\theta_i$, which is measured in radians and chosen at random from the interval [0, $2\pi$]. The update rules~\cite{Vicsek95} at time $t \neq 0$ for the position and the orientation of $i$th individual have the form
\begin{eqnarray}
\label{eq:e1}
x_i(t+\Delta t) &=& x_i(t) + v_0\Delta t\cos\theta_i(t),\\
\label{eq:e2}
y_i(t+\Delta t) &=& y_i(t) + v_0\Delta t\sin\theta_i(t),\\
\label{eq:e3}
\theta_i(t+\Delta t) &=& \overline{\theta_i(t)}_{S(i)}+\eta,
\end{eqnarray}
where $\Delta t$ is a time step, $x_i(t)$, $y_i(t)$ and $\theta_i(t)$ are the components of the position and the orientation of $i$th individual at time $t$, respectively. The last term in Eq.~(\ref{eq:e3}) is a random number generated from the uniform distribution on an interval $[-\eta/2, \eta/2]$, $\eta$ is so called  ``noise''. The notation $\overline{\theta_i(t)}_{S(i)}$ is the average direction of the velocity of individuals within an fixed area $S(i)$, which is defined by the angle
\begin{equation}
\label{eq:theta}
\overline{\theta_i(t)}_{S(i)}=\arctan\left(\frac{\sum_{j\in S(i)}\sin\theta_j(t)}{\sum_{j\in S(i)}\cos\theta_j(t)}\right).
\end{equation}

In biology, the eye structure of animal species is very different, and can be  classified into two main groups: (i) Animals with their eyes at the front of their head are the case of most of predators, omnivores or carnivores (for example cats, dogs, foxes, wolves and the weasel family…). Their eye structure gives them a binocular vision which enables them to focus and see quick movements from far away distances. In addition, they also can  easily contract or dilate their eyes's pupils depending on whether the light is bright or dim. (ii) Animals with their eyes on the side of their head are the case of most of preys; the structure of their eyes gives them the peripheral vision, an advantage to sense dangers such as predators. Of course, animals can be both predator and prey which depend on the situation they are facing, for example, a small fish can be a predator to a zooplankton but it can be also a prey for a bigger fish.

\begin{figure}[t]
\centering
\includegraphics[width=1.2in]{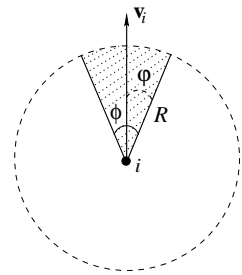}%
\caption{The angle of view of individual $i$, denote $\varphi = \phi/2$ is a half of vision angle.\label{fig:f1}}
\end{figure}

Based on the above biological property, in this work we assume the direction of motion of individual $i$ depends on the average orientation of the individuals (including the $i$th itself) within a circle sector $S(i)$ centered on $i$ with central angle $\phi$ and radius $R$ (see Fig.~\ref{fig:f1}). For simplicity, we use another notation for the central angle $\varphi = \phi/2$. In the case $\varphi = \pi$ (i.e. $\phi = 2\pi$), we obtain the original VM.

If a group of animals of the same species are synchronizing their motion so they all move in the same direction, then the animal are said to be flocking. In the opposite limit, they run away with  a maximum speed from a danger such as a predator. The change of the order to the disorder limits is similar to the phase transition of a ferromagnetic spin system, in which the phase change from ordered to disordered phase occurs at the transition temperature. The ordered or disordered phases correspond to the flocking or running away behavior of animals, and the noise corresponds to the temperature.

Now we consider the velocity $\bm v_i$ of $i$th individual as a spin vector $\bm S_i$ in the classical XY model, where the order parameter is defined by the absolute value of the average normalized velocity~\cite{Vicsek95}:
\begin{equation}
\label{eq:order}
Q = \frac{1}{N v_0}\left|\sum_{i=1}^N \bm v_i\right|,
\end{equation}
in which, $N$ is the total number of individuals. Of course, the order parameter is a function of the noise $\eta$ (in Eq.~\ref{eq:e3}). The variance of order parameter is written in the form
\begin{equation}
\label{eq:sus}
\sigma = \langle Q^2\rangle -\langle Q\rangle^2.
\end{equation}
Using this quantity, we can obtain the critical value of noise $\eta_c$ determined at the maximum of $\sigma$.

\section{Simulation results}
For all the simulations, we used the radius of circle sector $R=1$ and time step $\Delta t=1$. The absolute velocity $v_0 = 0.03$ and a fixed density $\rho = N/L^2 \simeq 1$, where the plane size $L = \mathrm{int}(\sqrt{N})$ with the total number of individuals $N = 40,\ 100,\ 200,\ 300,\ 400$ and $500$, $N$ is called the system size. The angle of view $\phi$ is varied from $0$ to $2\pi$, i.e. $\varphi \in [0,\pi]$.
\begin{figure}[t]
\centering
\includegraphics[width=2.7in]{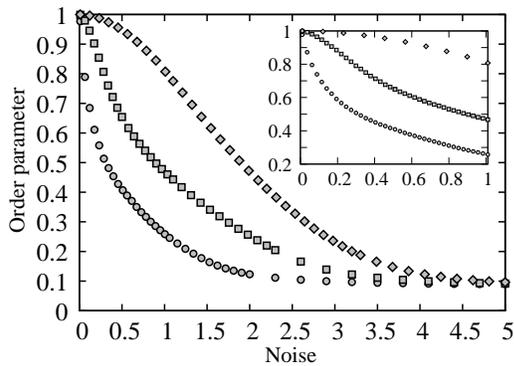}%
\caption{Order parameter versus noise $\eta$ for values of $\varphi$: $0.1\pi$ (circles), $0.5\pi$ (squares) and $1.0\pi$ (diamonds), with the system size $N=100$. The inset shows the enlarged scale.\label{fig:f2}}
\end{figure}

\begin{figure}[t]
\centering
\includegraphics[width=2.7in]{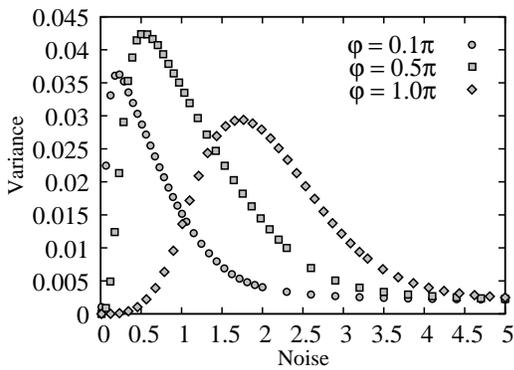}%
\caption{Variance of the order parameter versus noise $\eta$ for values of $\varphi$: $0.1\pi$ (circles), $0.5\pi$ (squares) and $1.0\pi$ (diamonds), with the system size $N=100$.\label{fig:f3}}
\end{figure}

Figure \ref{fig:f2} shows the dependence of the order parameter $\langle Q\rangle$ on the noise $\eta$ for vision angles $\varphi = 0.1\pi$ (circles), $\varphi= 0.5\pi$ (squares) and $\varphi = 1.0\pi$ (diamonds), and the system size $N=100$. Particularly, at $\varphi = \pi$, the order parameter is in good agreement with the results of previous simulations~\cite{Vicsek95}. These curves have the same shape as the magnetization in the spin model.
At low noise, most of the individuals move in the same direction at a constant speed $v_0$, and the order parameter $\langle Q\rangle \rightarrow 1$. Thus, the phase is an ``ordered" phase. Fluctuations of the orientation of the individuals  increase with increasing noise. Then, the order parameter tends to zero at high noise where all the individuals have random orientations, this phase is the ``disordered" phase.

\begin{figure}[t]
\centering
\includegraphics[width=2.8in]{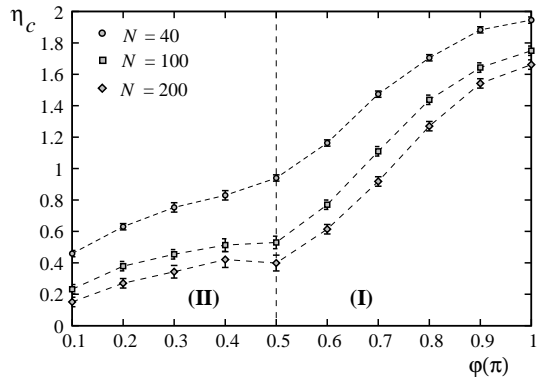}%
\caption{The phase diagram in the space ($\varphi$, $\eta_c$) with the system sizes $N=40,\ 100,\ 200$.\label{fig:f4}}
\end{figure}

\begin{figure}[t]
\centering
\includegraphics[width=2.5in]{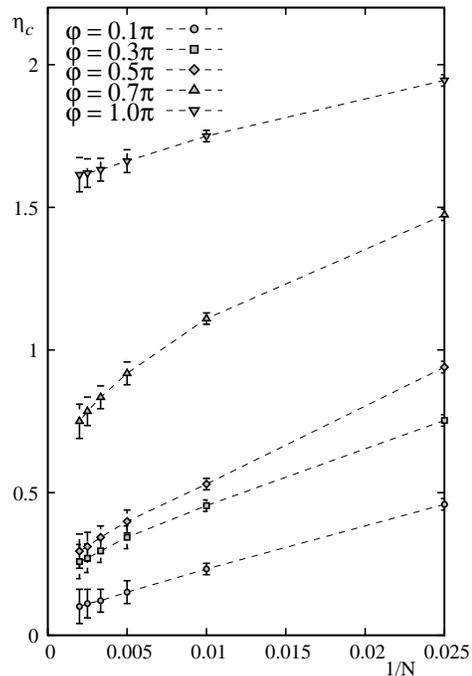}%
\caption{The critical noise $\eta_c$ versus $1/N$ for several values of $\varphi$: $0.1\pi$, $0.3\pi$, $0.5\pi$, $0.7\pi$ and $1.0\pi$.\label{fig:f5}}
\end{figure}

As shown in Fig.~\ref{fig:f3}, the critical value of noise $\eta_c$ is obtained by the value of $\eta$ at the maximum of the variance $\sigma$, one has $\eta_c = 0.232,\ 0.53$ and $1.75$ for $\varphi= 0.1\pi,\ 0.5\pi$ and $1.0\pi$, respectively. The ordered phase corresponds to a low noise $\eta <\eta_c$, and the disordered phase to a high noise $\eta > \eta_c$.

The phase diagram in the space ($\varphi$, $\eta_c$) has been shown in Fig.~\ref{fig:f4}, where $\varphi = \phi/2$ in units of $\pi$. The critical noises increase with increasing the vision angle, it indicates that more angle of view is better for flocking behavior of animals. These are contrary to the results of Gao {\it et. al.,} where the critical noise increases with decreasing the restricted angle~\cite{Jianxi}. We also see that the phase transition has been separated into two kinds at $\varphi = 0.5\pi$: phase (I) corresponds to the vision angle of the prey which is about 360 (degree), and phase (II) corresponds to the one of the predator ($\phi < 180 \ \mathrm{(degree)}$). The separation is much clearer with increasing system size.
For the vision angle $\varphi < 0.5\pi$, the critical noise $\eta_c$ is very small, this is perhaps a frozen state at low temperature in the physical systems, but not a phase transition.

We shown in Fig. 5 the critical noise decreases with increasing the system size $N$. This is a signature of a first-order transition~\cite{ngo08,ngo10}, but the order parameter has not a discontinuity at $\eta_c$. For $\varphi > 0.5\pi$, $\eta_c$ converges to a constant with $N\rightarrow \infty$, while it tends to zero for $\varphi < 0.5\pi$. It indicates that there is no phase transition in the flocking behavior of a dense group of predators.

\section{Conclusions}
We studied the effects of the vision angle on the phase transition behavior in systems of animals. This model is equivalent to a ferromagnetic XY spin system in which, we align each spin (i. e. we move an animal) along the local field coming from moving directions of animals by adding some noise. In the case of short-range interaction, the XY spins in two-dimensions has no long-range ordering at finite noise. It is the well-known Kosterlitz-Thouless transition. In the present model, the local field results from a long-range interaction, in addition to an orientational anisotropy $\varphi$, which explains the existence of a long-range ordering at finite noise.
Our simulation results show the critical noise $\eta_c$ quickly decreased with decreasing angle of view $\varphi$. The phase transition strongly depends on the vision angle of each individual $\varphi$ in range of $(\pi/2,\pi)$. This range of vision angle corresponds to the structure and function of eyes of the prey (the viewable area $\phi \in [0,2 \pi]$ or $\varphi >\pi/2$), but not the ones of the predator $(\phi \in [0,\pi])$. So, we can conclude that the flock of animals is a common behavior of prey species. This behavior sometimes occurs to the group of predator when they are in the face of danger.

\acknowledgments
The authors are grateful to H. T. Diep for helpful comments and suggestions. This work was supported by the Nafosted (Vietnam National Foundation for Science and Technology Development), Grant No. 103.02-2011.55.

\end{document}